\documentclass[conference]{IEEEtran}
\usepackage{float}
\usepackage{cite}
\usepackage{amsmath,amssymb,amsfonts}
\usepackage{graphicx}
\usepackage{textcomp}
\usepackage{xcolor}
\usepackage{algpseudocode}
\usepackage{comment}

\usepackage{algorithm}
\usepackage{amsmath}  
\usepackage{microtype}
\usepackage{adjustbox}



\def\BibTeX{{\rm B\kern-.05em{\sc i\kern-.025em b}\kern-.08em
    T\kern-.1667em\lower.7ex\hbox{E}\kern-.125emX}}
\begin{document}

\title{Leveraging Caliper and Benchpark to Analyze MPI Communication Patterns: Insights from AMG2023, Kripke, and Laghos\\

}

\author{\IEEEauthorblockN{1\textsuperscript{st} Grace Nansamba}
\IEEEauthorblockA{\textit{Department of Computer Science)} \\
\textit{Tennessee Tech University.}\\
Tennesse, USA \\
gnansamba42@tntech.edu}
\and
\IEEEauthorblockN{2\textsuperscript{nd} Evelyn Namugwanya}
\IEEEauthorblockA{\textit{Department of Computer Science} \\
\textit{Tennessee Tech University}\\
Tennesse, USA \\
enamugwan42@tntech.edu}
\and
\IEEEauthorblockN{3\textsuperscript{rd}David Boehme}
\IEEEauthorblockA{\textit{Center for Applied Scientific Computing} 
\\
\textit{Lawrence Livermore National laboratory}\\
California, USA \\
boehme3@llnl.gov}
\and
\IEEEauthorblockN{4\textsuperscript{th}Dewi Yokelson}
\IEEEauthorblockA{\textit{Center for Applied Scientific Computing} 
\\
\textit{Lawrence Livermore National laboratory}\\
California, USA \\
yokelson1@llnl.gov}
\and
\IEEEauthorblockN{5\textsuperscript{th} Riley Shipley}
\IEEEauthorblockA{\textit{Department of Computer Science} \\
\textit{Tennessee Tech University.}\\
Tennesse, USA \\
rshipley@tntech.edu}
\and
\IEEEauthorblockN{6\textsuperscript{th}Derek Schafer}
\IEEEauthorblockA{\textit{Department of Computer Science} 
\\
\textit{University of New Mexico}\\
New Mexico, USA \\
dschafer1@unm.edu}
\and
\IEEEauthorblockN{7\textsuperscript{th}Michael McKinsey}
\IEEEauthorblockA{\textit{Center for Applied Scientific Computing} 
\\
\textit{Lawrence Livermore National laboratory}\\
California, USA \\
mckinsey1@llnl.gov}
\and
\IEEEauthorblockN{8\textsuperscript{th}Olga Pearce}
\IEEEauthorblockA{\textit{Center for Applied Scientific Computing} 
\\
\textit{Lawrence Livermore National laboratory}\\
California, USA \\
pearce8@llnl.gov}
\and
\IEEEauthorblockN{9\textsuperscript{th} Anthony Skjellum}
\IEEEauthorblockA{\textit{Department of Computer Science} \\
\textit{Tennessee Tech University.}\\
Tennesse, USA \\
askjellum@tntech.edu}
}

\maketitle
\begin{abstract}
We introduce  ``communication regions'' into the widely used Caliper HPC profiling tool. 
A communication region is an annotation  enabling capture of metrics about the data being communicated (including statistics of these metrics), and metrics about the MPI processes involved in the communications, something not previously possible in Caliper.
We explore the utility of communication regions with three representative modeling and simulation applications---AMG2023, Kripke, and Laghos---all part of the comprehensive Benchpark suite that includes Caliper annotations.  Enhanced Caliper  reveals 
detailed communication behaviors. 
Using Caliper and Thicket in tandem, we   create new visualizations of  MPI communication patterns, including halo exchanges. 
Our findings reveal communication bottlenecks and detailed behaviors,
%
indicating significant utility of the special-regions addition to Caliper.  The comparative scaling behavior of both CPU- and GPU-oriented systems are  shown; we are able to look at different regions within a given application, and see how scalability and message-traffic metrics differ.
%
\end{abstract}


\section{Introduction}
In high-performance computing (HPC), the Message Passing Interface (MPI)\cite{mpi41} is widely used for communication between processes in parallel applications. Detecting and optimizing communication patterns are crucial to achieving optimal performance, particularly as systems scale. They can differ markedly from single built-in collective operations in MPI to composites of point-to-point and collective operations 
that comprise application-specific data movement.  As the complexity of applications and hardware grows, understanding and optimizing these patterns has become increasingly important for performance tuning and resource management.
The performance behavior of an MPI application depends on communication and computation costs. Application developers strive to minimize communication costs so that most of the application's time is spent on the main  computation rather than communication and other overheads. Analyzing communication patterns in MPI application is fundamental to improving their design and overall performance. 

This work explores MPI communication pattern analysis from two complementary perspectives.
First, we introduce a new feature to the Caliper~\cite{Caliper} profiling library to compute MPI communication statistics for special communication regions. These regions mark groups of MPI calls that form logical communication phases, such as halo exchanges. By applying this approach to benchmark applications such as Kripke~\cite{Kripke} and AMG2023~\cite{AMG2023}, we are able to visualize how communication patterns vary across different configurations (scales and data layouts), revealing  insights into communication overheads, load balancing, and scalability.
Second, we utilize Benchpark~\cite{Benchpark2024}, a collaborative repository for reproducible specifications of HPC benchmarks, and Thicket~\cite{Thicket2023}, a Python-based toolkit for exploratory data analysis of parallel performance data, to automate the data collection as well as analyses of scaling studies for three benchmarks considered here.

The Benchpark suite provides a comprehensive collection of benchmarks, including AMG2023, Kripke, and Laghos~\cite{LAGHOS_REAL}, where communication patterns such as halo exchanges and communication regions such as MatVecComm and sweep 
play a critical role in performance. By profiling these applications, we aim to identify patterns and operational configurations that have communication bottlenecks.

The contributions of this paper are as follows:
\begin{itemize}

\item Extended Caliper's instrumentation interface with special markers for communication patterns (communication regions) and a communication pattern profiler that records communication statistics such as message sizes, sources, and destinations.

\item Manually annotated communication regions to three applications (AMG2023, Kripke, and Laghos) in order to capture performance metrics of key communication phases.

\item Conducted experiments using Benchpark to build these three benchmarks with Caliper measurement support and executed them with a range of scaling configurations.



\item Showed comparative scaling behavior across the three benchmark applications on both CPU- and GPU-oriented systems.

\end{itemize}

The remainder of this paper is organized as follows: Section~\ref{sec:background} presents  background information, while 
Section~\ref{sec:methodology} details the methodology and introduces the new communication patterns we integrated into Caliper. Sections~\ref{sec:results} and \ref{sec:bandwidth} jointly report  scaling results and highlights key insights derived from each benchmark, and Section~\ref{sec:conclusion} concludes the paper by summarizing the contributions and observations.

\section{Background}
\label{sec:background}

Caliper~\cite{Caliper} is an instrumentation and performance profiling library for HPC codes. 
It features a source-code instrumentation API to mark regions of interest in a HPC program, as well as built-in analysis functionality for HPC programming models including MPI, OpenMP, CUDA, and HIP. For MPI, Caliper intercepts API calls via either PMPI or GOTCHA~\cite{GOTCHA}, a runtime function-wrapping library, and inspects their parameters to record statistics. 

Performance profiling tools that have overlapping funcionality with Caliper include TAU~\cite{Shende2006TheTP, Malony2011AdvancesIT}, Vampir\cite{Vampir2001}, and HPCToolkit~\cite{Adhianto2010HPCTOOLKITTF}. These  time function calls including MPI calls with either instrumentation or sampling methods and then generate profiles of performance data. In addition to time spent in functions, they can gather other metrics of interest, such as memory allocations.

Thicket~\cite{Thicket2023} provides exploratory data analysis capabilities for any application that has been profiled by Caliper. 
Caliper performance profiles are easily uploaded into Thicket objects that can be easily manipulated with Python data analysis and visualization libraries to generate statistics and plots for understanding performance. 
The Benchpark framework~\cite{Benchpark2024} enables automated continuous benchmarking across heterogeneous HPC systems. With reproducible specifications for various systems and benchmarks, scaling experiments can be streamlined across new architectures. 

To better understand and optimize communication in irregular HPC applications, Hoefler et al. introduced scalable communication protocols for dynamic sparse data exchange (DSDE)~\cite{Hoefler}, a challenge that arises in modern large-scale applications with non-uniform communication patterns. Their work highlights the limitations of traditional bulk synchronous parallel (BSP) models, especially when applied to dynamic and sparse communication phases. To address this, they proposed the NBX algorithm, a lightweight protocol designed to reduce memory overhead and improve efficiency in sparse communication scenarios. This contribution emphasizes the need for protocol-level advancements to support increasingly irregular and dynamic application behavior, an objective that resonates with our efforts to expose and analyze communication patterns through communication regions in Caliper and Benchpark.

The increasing adoption of software-based virtualization in cloud environments has introduced new challenges for high-performance computing (HPC) applications, particularly in communication efficiency. Shao et al.~\cite{Shao} analyzed the communication performance of MPI applications in virtualized systems and observed significant performance degradation when virtual CPUs (vCPUs) become overcommitted on a physical host. Their work identifies a critical yet underexplored bottleneck in HPC communication, emphasizing that virtualization layers can exacerbate latency and interrupt MPI progress. By investigating the root causes of this degradation and proposing corrective strategies, their study provides valuable insight into how communication patterns and system-level configurations can influence MPI behavior. This context reinforces the importance of communication-aware profiling, such as our approach using communication regions in Caliper, especially when targeting modern and potentially virtualized environments.

\section{Methodology}
\label{sec:methodology}


The communication characteristics of a parallel applications are generally determined by their underlying logical communication pattern(s). In MPI, some communication patterns can be expressed as a single collective operation, but localized and/or sparse communication patterns are usually implemented through several individual point-to-point operations. Since communication patterns are logical concepts, reasoning about communication patterns based on regular application performance profiles or traces can be difficult. To address this gap, we added two extensions to the Caliper performance profiling library: (1) a set of special instrumentation markers to mark code regions that form a communication pattern, and (2) a communication pattern profiler that computes communication statistics for each communication pattern instance.


In this work, we provide two new markers for applications to denote the start and end of a specific communication region: \texttt{CALI\_MARK\_COMM\_REGION\_BEGIN} and \texttt{CALI\_MARK\_COMM\_REGION\_END}. These markers can be placed around groups of MPI calls which form a logical communication pattern in a program, and be given a name for later reference. For example, all point-to-point operations that form a halo exchange in a stencil-based domain decomposition can be logically grouped together this way. These markers let us distinguish MPI operations belonging to one communication pattern instance from operations in a different communication pattern or another instance of the same pattern in a subsequent iteration. The communication pattern profiler is invoked at the end of each marked communication region and computes message, rank, and data volume statistics for the MPI operations that occurred within the region boundaries. Table~\ref{tab:attributes}
lists the attributes that the communication pattern profiler can collect.

\begin{table}
\centering
\caption{MPI Attributes Collected by Caliper\label{tab:attributes}}
\renewcommand{\arraystretch}{1.25}  
\setlength{\tabcolsep}{8pt} 
\begin{tabular}{|l|l|}
\hline
\textbf{Attribute}       & \textbf{Description}                       \\
\hline
Sends & Min/Max\hbox{.} number of messages sent \\
\hline
Recvs & Min/Max\hbox{.} number of messages received \\
\hline
Dest ranks & Min/Max\hbox{.} number of distinct destination ranks \\
\hline
Src ranks & Min/Max\hbox{.} number of distinct source ranks \\
\hline
Bytes sent & Min/Max\hbox{.} message size sent by a process in a region \\
\hline
Bytes recv & Min/Max\hbox{.} message size received by a process in a region \\
\hline
Coll & Max\hbox{.} collective calls in a region \\
\hline
\end{tabular}
\end{table}

\subsection{Benchmarks Overview}
\label{sec:benchmarks}
This study evaluates communication patterns and scalability using three key benchmarks: {AMG2023}, {Kripke}, and {Laghos}. All three are run using Benchpark reproducible specifications for strong and weak scaling experiments. Each benchmark represents a distinct computational model and communication behavior:

\begin{itemize}
    \item \textbf{AMG2023}: An Algebraic Multigrid solver designed to solve large sparse linear systems derived from discretized partial differential equations~\cite{AMG2023}. AMG2023 exhibits distributed communication patterns that reflect its global dependencies. AMG2023 depends on \textit{hypre}\cite{hypre}.
    \item \textbf{Kripke}: A particle transport mini-app that explores data layouts and programming approaches for deterministic Sn transport problems in three dimensions~\cite{Kripke}. Its communication is highly localized, with limited dependencies on neighboring processes.
    \item \textbf{Laghos}: A high-order Lagrangian hydrodynamics solver for compressible gas dynamics~\cite{LAGHOS_REAL}. Laghos involves intricate communication patterns because of its dynamic mesh deformation and high-order finite element computations. Laghos depends on MFEM~\cite{MFEM}.
\end{itemize}

\subsection{MPI Regions of Interest}
\label{region_o_interest}

All three benchmarks analyzed in this study (AMG2023, Kripke, and Laghos) contain at least one \textbf{halo exchange region}, a critical communication pattern in distributed high-performance computing. A \textbf{halo exchange} refers to the exchange of boundary data between neighboring processes in a domain-decomposed parallel application. This is essential in scenarios where processes require information from adjacent regions to perform computations accurately, such as updating grid points in numerical simulations or solving partial differential equations. Halo exchanges are characterized by localized point-to-point communication.

The benchmarks employ halo exchanges because they align with the computational frameworks of these applications. For example, AMG2023~\cite{AMG2023} uses halo exchanges to ensure consistency across subdomains in its multigrid hierarchy, Kripke ~\cite{Kripke} leverages them to synchronize particle transport data between neighboring domains, and Laghos~\cite{LAGHOS_REAL} relies on them for updating mesh boundaries in its dynamic finite element framework. Measuring halo exchanges (aka ghost-cell exchanges) 
is particularly important as they often dominate the communication overhead in applications. By profiling halo exchanges, we can identify bottlenecks, assess load balancing, and evaluate scalability, thus offering opportunities for optimization.

The \textbf{Sweep region} computes the transport solution over a subdomain by traversing cells in dependency order, applying upwind conditions, solving the local equation, and communicating results.

The \textbf{MatVecComm region} sets up the MPI communication structure need\-ed to perform a distributed matrix-vector product in \textit{hypre}, particularly by defining how to exchange off-rank vector values efficiently.


\subsection{Experimental Setup}
\label{sec:experiments}
To evaluate these benchmarks, experiments were conducted on two HPC architectures
described in Table~\ref{tab:machines}.

\begin{table}[htbp]
\centering
\renewcommand{\arraystretch}{1.25}  
\setlength{\tabcolsep}{8pt} 
\small 
\caption{Architectures used for the experiments \label{tab:machines}}
\begin{adjustbox}{max width=\columnwidth}
\begin{tabular}{|l|c|c|}
        \hline
        \textbf{Hardware Attribute} & \textbf{Tioga} & \textbf{Dane} \\ 
        \hline
        \textbf{CPU Architecture}  & AMD Trento & Intel Sapphire Rapids \\
        \hline
        \textbf{CPU Cores / Node} & 64 & 112 \\
        \hline  
        \textbf{Memory (GB) / Node}  & 512 & 256 \\
        \hline
        \textbf{GPU Architecture}  & AMD MI250X & N/A \\
        \hline
        \textbf{\# GPUs / Node}  & 8 & N/A \\
        \hline
\end{tabular}
\end{adjustbox}
\end{table}

The experiments were designed to capture mainly \textbf{weak scaling} (problem size increases proportionally with processes) because scientists often aim to solve larger, more detailed problems rather than merely accelerating smaller ones. For comparison purposes, the same 3-dimensional problem sizes and MPI ranks were used for both AMG2023 and Kripke. We also captured \textbf{strong scaling} (fixed problem size, increasing number of processes) for Laghos (Lagrangian High-Order Solver). Details for the experimental configurations are shown in Table~\ref{tab:experiments_combined}. 
Key metrics such as maximum bytes sent, source and destination ranks were analyzed to identify trends in communication patterns.
 Table~\ref{tab:bytesAll} shows a handful of these metrics from the annotated regions in the selected benchmarks.

\subsection{Benchpark design and use }
Benchpark\cite{Benchpark2024} uses Spack and Ramble to build benchmarks and configure applications across different systems, using modifiers to encapsulate reusable patterns for a given experiment. Each experiment defines a particular benchmark, such as Kripke, with defined configurations which are reproducible across different systems/machines. 
The Caliper modifier enables profiling in Benchpark and has different variants which are defined at experimental setup such as \texttt{mpi} and \texttt{cuda}. The new MPI attributes collected by Caliper were added to this modifier. 

\begin{table}[htbp]
    \centering
    \caption{
    Experiments run for each benchmark on the Dane and Tioga systems}  
    \scriptsize
    \setlength{\tabcolsep}{2pt}
    \renewcommand{\arraystretch}{1.25}
    \begin{adjustbox}{max width=\columnwidth}

    \begin{tabular}{|c|c|c|c|c|c|}
    \hline
      \textbf{Benchmark} & \textbf{Scaling Type}&  \textbf{Smallest Problem} & \textbf{\# Nodes} & \textbf{\# Processes} & \textbf{Dimensions} \\ 
      \hline
       \multicolumn{6}{c}{\textbf{Dane}} \\
    \hline
     AMG2023 
     & Weak &   32$\times$32$\times$16 & 1,2,3,5 & 64,128,256,512 & 4$\times$4$\times$4,8$\times$4$\times$4,\\ 
    &&&&& 8$\times$8$\times$4,8$\times$8$\times$8 \\ 
      
    \hline
    Kripke 
     & Weak  & 16$\times$32$\times$32 & 1,2,3,5 & 64,128,256,512 & 4$\times$4$\times$4,8$\times$4$\times$4,\\
     &&&&& 8$\times$8$\times$4,8$\times$8$\times$8 \\ 
    \hline
    Laghos &Strong & & 1,2,4,8 & 112,224,448,896 & \\ 
    \hline
    \multicolumn{6}{c}{\textbf{Tioga}} \\ 
    \hline
        AMG2023 
     & Weak & 32$\times$32$\times$16 & 1,2,4,8 & 8,16,32,64& 2$\times$2$\times$2,4$\times$2$\times$2,\\ 
     &&&&& 4$\times$4$\times$2,4$\times$4$\times$4  \\ 
    \hline
    Kripke 
     & Weak   & 16$\times$32$\times$32 & 1,2,4,8 & 8,16,32,64 & 2$\times$2$\times$2,4$\times$2$\times$2, \\
     &&&&& 4$\times$4$\times$2,4$\times$4$\times$4 \\ 
    \hline
    \end{tabular}
     \end{adjustbox}
    \vspace{-0.25cm}
    \label{tab:experiments_combined}
\end{table}

\begin{table}[ht]
    \centering
    \caption{Sample metric collection from annotated application regions}
    \begin{adjustbox}{max width=\columnwidth}
    \begin{tabular}{|r|c|c|c|c|}
        \hline
        \textbf{Application and} & \textbf{Total} & \textbf{Total} & \textbf{Largest} & \textbf{Average} \\
        \textbf{Number of}& \textbf{Number of} &  \textbf{Number} & \textbf{Send} & \textbf{Send Size} \\
        \textbf{Processes} & \textbf{Bytes Sent} & \textbf{of Sends} & \textbf{(bytes)} &  \textbf{(bytes)} \\
        \hline
        Laghos - 112 & 3.20E+11 & 2.55E+08 & 80256 & 1.26E+03 \\
        224 & 3.71E+11 & 5.02E+08 & 51792 & 7.39E+02 \\
        448 & 4.26E+11 & 9.35E+08 & 33824 & 4.55E+02 \\
        896 & 5.31E+11 & 1.79E+09 & 29072 & 2.96E+02 \\
        \hline
        Kripke (Dane) - 64  & 4.03E+09 & 184320  & 8388608 & 2.18E+04 \\
        128 & 8.72E+09 & 389120  & 8388608 & 2.24E+04 \\
        256 & 1.81E+10 & 819200  & 8388608 & 2.21E+04 \\
        512 & 3.76E+10 & 1720320 & 8388608 & 2.18E+04 \\
        \hline
        Kripke (Tioga) - 8   & 1.34E+09 & 15360   & 16777216 & 8.74E+04 \\
        16  & 3.36E+09 & 35840   & 25165824 & 9.36E+04 \\
        32  & 7.38E+09 & 81920   & 29360128 & 9.01E+04 \\
        64  & 1.61E+10 & 184320  & 33554432 & 8.74E+04 \\
        \hline
        AMG2023 (Dane) - 64  & 1.98E+08 & 425880  & 35392    & 4.66E+02 \\
        128 & 6.83E+08 & 1016184 & 60992    & 6.72E+02 \\
        256 & 2.32E+09 & 2502084 & 102976   & 9.26E+02 \\
        512 & 6.96E+09 & 7735364 & 136256   & 8.99E+02 \\
        \hline
        AMG2023 (Tioga) - 8   & 1.56E+07  & 22080   & 17032    & 705.46 \\
        16  & 5.79E+07  & 65600   & 34604    & 881.97 \\
        32  & 2.14E+08 & 192720  & 68128    & 1110.27 \\
        64  & 7.28E+08 & 572520  & 136256   & 1271.07 \\
        \hline
    \end{tabular}
     \end{adjustbox}
    \label{tab:bytesAll}
\end{table}

\section{Scaling Results Per Benchmark}
\label{sec:results}


In this section, we evaluate the scaling behavior of the AMG2023, Kripke, and Lahgos benchmarks to 
observe the characteristics of the unique communication patterns in these HPC applications, which in turn can help reveal communication bottlenecks and scalability challenges. 

Table~\ref{tab:bytesAll}
includes the maximum number of bytes sent across various process counts for Kripke, AMG2023, and Laghos on Dane and Tioga. Kripke shows constant communication per rank on Dane and increasing volume on Tioga, reflecting Kripke's localized patterns and GPU bandwidth advantage. AMG2023 exhibits steadily increasing communication across both systems, consistent with its multigrid hierarchy. Laghos shows decreasing bytes sent with scale under strong scaling, highlighting reduced data per rank. 

\subsection{Kripke}
\label{sec:Kripke}
Using Kripke, we investigated point-to-point communication patterns on both Dane and Tioga. 
We annotated and analyzed the \textit{sweep} region, a regular communication pattern in a 3D cartesian grid to exchange halo data between neighboring subdomains. 
The number of communication partners (dest/source ranks) for each rank is either three or six, reflecting processes on the corner or in the middle of the domain. For the smallest GPU run every rank has only three communication partners because all ranks are on a corner. 
Every rank sends 36 messages to each rank in each communication phase. 

\begin{figure*}
    \centering
    \includegraphics[width=1.0\textwidth]{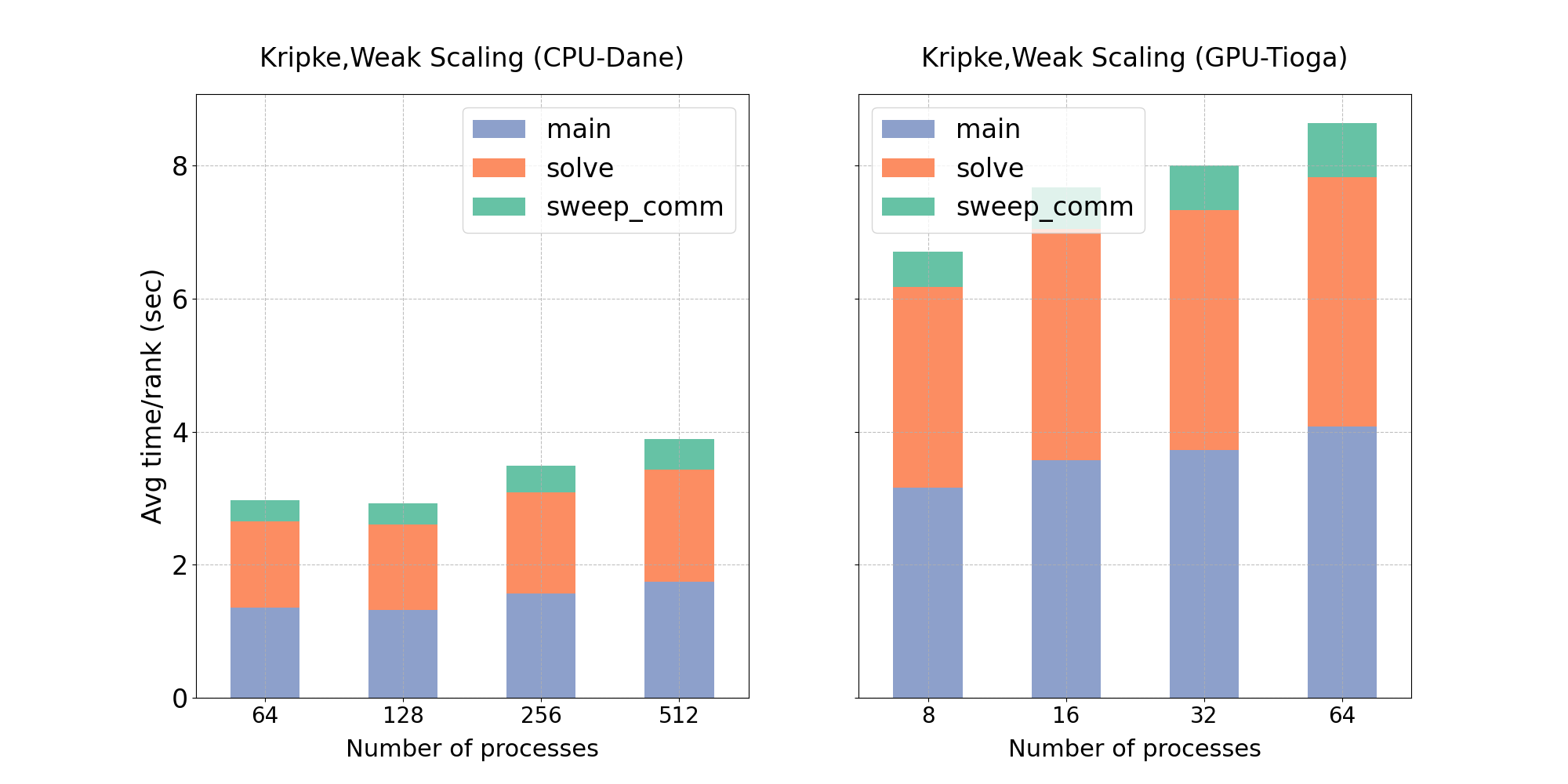}
    \caption{Average time per rank (seconds) for Kripke on Dane (left) and Tioga (right) }   
    \label{fig:kripeke_weakscaling_loops}
\end{figure*}

In Figure~\ref{fig:kripeke_weakscaling_loops}, the main loop represents the overall time for the application which is mainly comprised of the solve loop (computation) and \texttt{sweep\_comm} (communication). The relative time in \texttt{sweep\_comm} vs main loop is higher on Dane than on Tioga, showing that more time is taken in communication while using CPUs compared to GPUs. 
%
%
%
%
As problem size per rank increases under weak scaling conditions, the compute load grows, particularly in the \texttt{solve} loop. In contrast, \textit{sweep} communication remains relatively constant or grows slowly because halo exchanges are limited to neighbor processes and scale less aggressively. 

The \texttt{main} loop accounts for all application runtime overheads, including time spent in \texttt{solve}, \texttt{sweep\_comm}, and  operations such as initialization and synchronization. As the number of processes increases, platform-specific scaling behavior becomes evident. On GPUs, the \texttt{main} loop, \texttt{sweep\_comm} loop, and average \texttt{solve}  time increase as compared to the CPUs.

The observed trends are explained below.
\begin{itemize}
    \item \texttt{sweep\_comm} takes the least time because Kripke’s communication is light\-weight and often overlapped with computation.
    \item The \texttt{solve} loop dominates due to heavy arithmetic operations in transport sweeps.
    \item Performance and scaling behavior may depend on architecture bandwidth, parallel efficiency, and communication overheads.
    \item Optimizing Kripke may involve reducing \texttt{solve} cost on CPUs and ensuring the delay of the message is hidden in GPU runs.
\end{itemize}

\subsection{AMG2023}

{\color{red} 
}

We used AMG2023 to investigate how communication patterns (messages sent, bytes sent, src and dst ranks) behave across different multi-grid levels on both GPU and CPU based architectures. 
This section highlights how communication behavior varies with different AMG levels across these systems.

With each increase in scale, the number of MG levels goes up and more data needs to be communicated, meaning runs on Dane had more levels than those on Tioga. 
With each increase in level, the communication becomes increasingly non-local (number of dest/source ranks goes up) and at higher levels there are many more communication partners (over 100 in our case). Meanwhile most of the data volume is exchanged in more localized communication at the lower levels (bytes sent is much higher for the lower levels). This analysis is possible because we can distinguish the communication pattern for each level. 

\subsubsection{Communication Behaviors Across Multi-grid Levels of AMG2023}

\begin{figure*}
    \centering
    \includegraphics[width=1.0\textwidth]{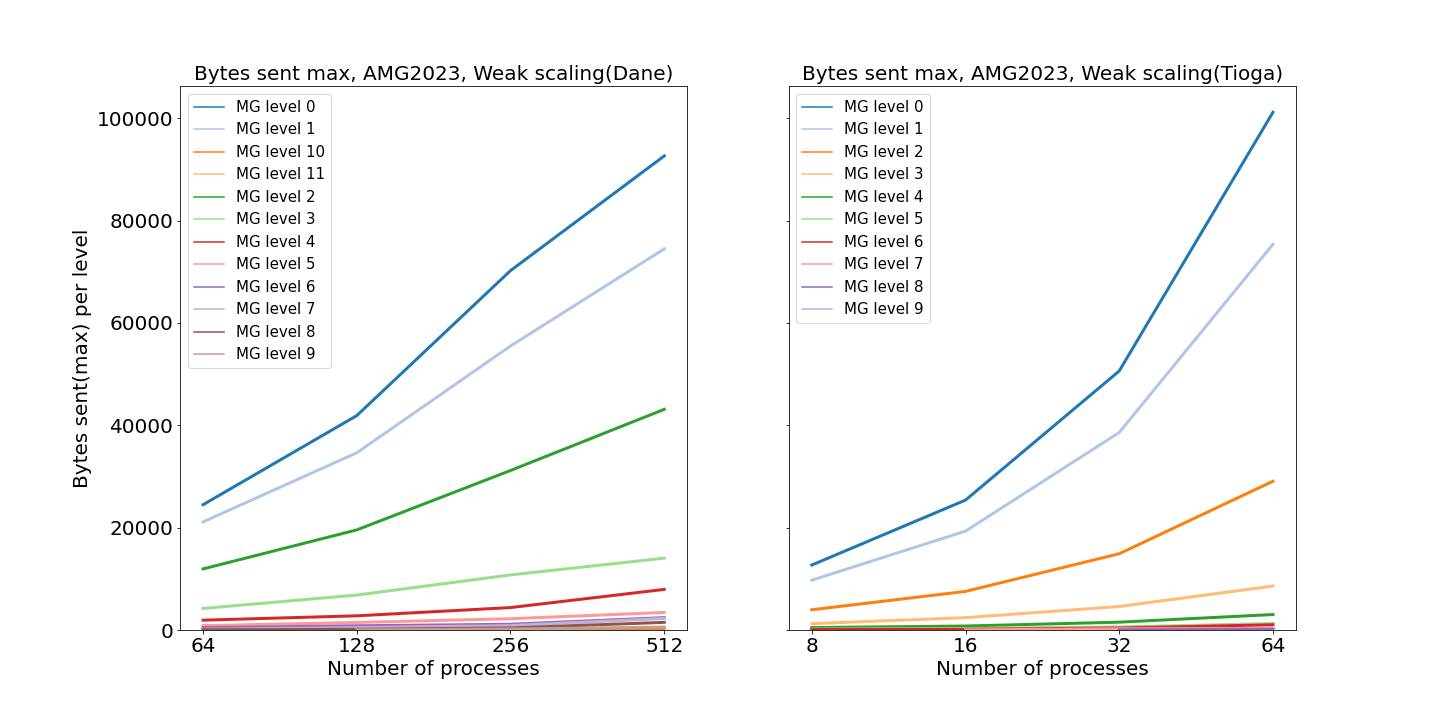}
    \caption{Bytes sent per level for AMG2023 on CPU-Dane(left) and GPU-Tioga(right) }   
    \label{fig:Bytes_per_secAMG_levels_weak_cpu_gpu}
\end{figure*}

Figure~\ref{fig:Bytes_per_secAMG_levels_weak_cpu_gpu} illustrates the maximum number of bytes sent per process across different multi-grid (MG) levels in AMG2023 under weak scaling. 
Each curve represents a distinct multigrid level, with level 0 denoting the finest grid and higher-numbered levels corresponding to progressively coarser grids.
\subsubsection{\color{black}Trend Observations }


Across both systems, we observe a clear variation in communication volume across the multigrid hierarchy:

\textbf{Fine Levels (MG level 0–2):} In Figure ~\ref{fig:Bytes_per_secAMG_levels_weak_cpu_gpu}, these levels exhibit the highest communication volumes, with bytes sent per process increasing significantly as the number of processes grows. This behavior is expected, as the finest levels dominate the computational workload and produce dense domain decompositions, which result in substantial inter-process communication due to the larger number of halo exchanges and boundary interactions.

\textbf{Intermediate Levels (MG level 3–5):} These levels show moderate communication growth, characterized by a noticeable but less increase in bytes sent compared to the finest levels. Although the problem is coarsened at these levels, inter-process communication remains non-negligible due to overlapping subdomains and the continued propagation of residuals and corrections between neighboring ranks.

\textbf{Coarse Levels (MG level $\geq$6):} The communication volume on the coarsest levels remains relatively flat or minimal throughout all the processes. At this stage, both computational and communication costs are negligible. This is due to the significantly reduced problem size, which often results in either collective communication or replicated coarse-level computations, minimizing the need for data exchange across ranks~\cite{hypre}.
\subsubsection{Comparison Between Architectures }
While the overall trends are consistent across the CPU and GPU platforms, the rate of increase in bytes sent varies.
On Dane, communication growth is more pronounced at both the fine and intermediate multigrid levels. This is likely due to the combination of higher per-message latency and limited memory bandwidth available on CPUs, which together amplify the cost of halo exchanges as the number of processes increases. As a result, the system communicates less efficiently at larger scales.

In contrast, the rate of communication growth on Tioga is more controlled. The presence of high-bandwidth memory and AMD’s Infinity Fabric interconnect enables more efficient data movement between processing elements. Additionally, GPU-optimized AMG implementations often exploit superior load balancing and data reuse strategies, particularly on coarse multigrid levels, which further suppress communication overhead and contribute to improved scaling performance.

\begin{figure*}
    \centering
    \includegraphics[width=1.0\textwidth]{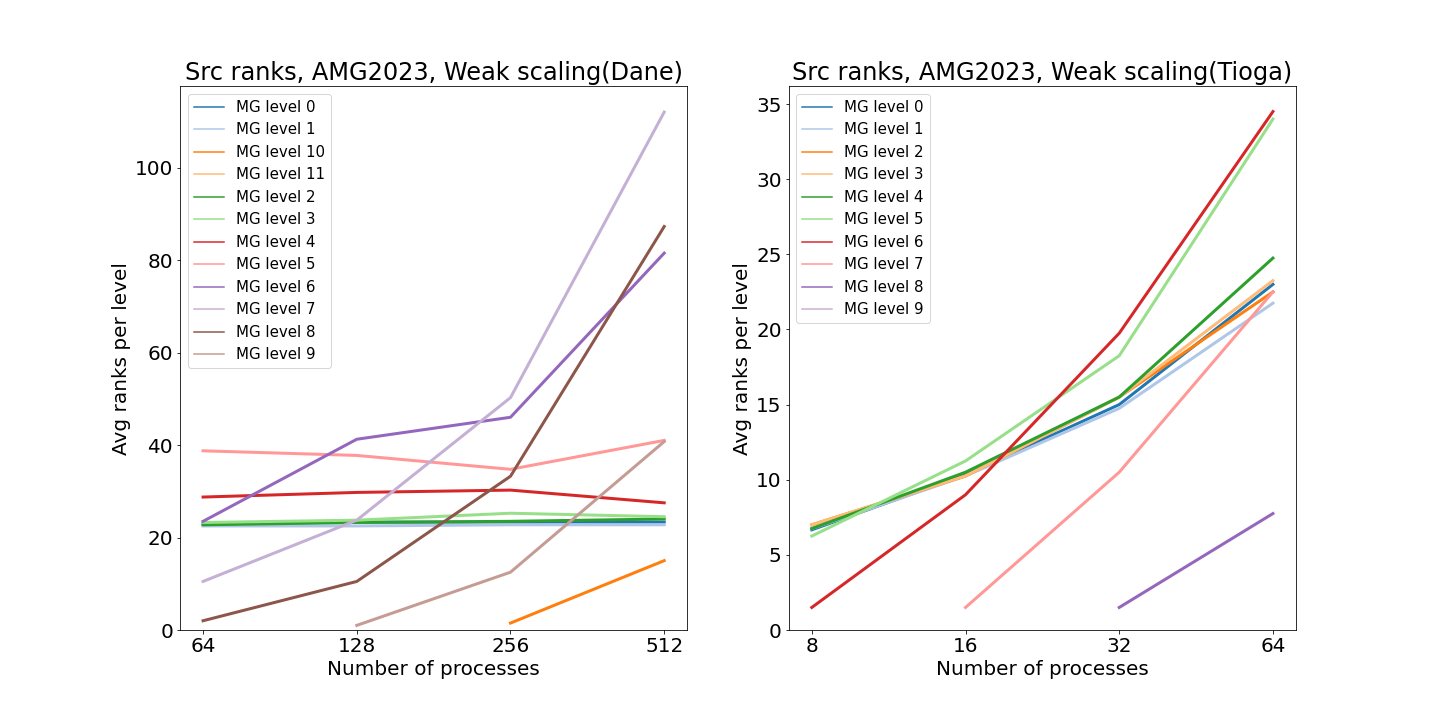}
    \caption{Average ranks per level for AMG2023 on Dane (left) and Tioga (right)  }
    \label{fig:amg_src_ranks}
\end{figure*}

\subsubsection{Source Rank Distribution Across Multigrid Levels in AMG2023}

Figure~\ref{fig:amg_src_ranks} presents the average number of source ranks per multigrid (MG) level in the AMG2023 benchmark under weak scaling conditions.
This metric reflects the number of MPI ranks actively participating in message sending for a given MG level, providing insight into the communication pattern and parallelism characteristics of AMG at scale.

\subsubsection{Trend Analysis on Dane} 

We observed different communication profiles across the hierarchy:

Low-numbered levels (MG levels 0--3) maintain a relatively flat number of source ranks across increasing process counts. This trend suggests that communication at finer resolutions is largely localized, and the participation of ranks remains stable due to consistent domain decomposition and limited need for global communication.

Mid- to high-numbered levels (MG levels 6--9), on the other hand, exhibit sharp increases in source rank participation, particularly beyond 256 processes. For example, MG level 6 involves more than 100 ranks for runs of 512 processes, indicating a significant broadening of the communication. This behavior points to increased communication complexity at coarse levels, which should ideally require less inter-process messaging.

This increase in the number of processes involved at higher MG levels may be attributed to inefficient coarse-level aggregation in the parallel AMG hierarchy, where suboptimal coarsening strategies cause the coarse problem to be distributed across more ranks than necessary, increasing communication overhead.
Such behavior can degrade scalability by increasing latency and load imbalance at coarse levels, which ideally should involve fewer participants due to the reduced problem size.

\subsubsection{Trend Analysis on Tioga }
In contrast, the GPU system exhibits a more gradual and uniform increase in source rank participation across all multigrid levels. Most MG levels (0--6) show nearly linear growth in the number of source ranks as the number of processes increases, indicating a well-balanced workload and a good distribution of communication responsibilities. Even at the coarser levels (MG 7--9), the number of participating ranks remains relatively low and grows at a slower rate than on the CPU-based platform.

These observations are consistent with the design objectives of GPU-optimiz\-ed AMG implementations, which typically incorporate load-balancing strategies across multigrid levels, efficient coarse-grid coarsening algorithms, and communication avoidance techniques. Such strategies are particularly important at the coarse levels, where GPUs may not be fully utilized due to limited concurrency.

The communication behavior observed on Tioga reflects the combined advantage of high-bandwidth memory (HBM2e), advanced interconnects like AMD’s Infinity Fabric, and AMG algorithmic enhancements that reduce unnecessary message traffic and ensure scalability.


\begin{itemize}
    \item \textbf{Finer levels} (MG levels 0-2) do not necessarily involve all processes uniformly across systems. On CPUs, this may reflect static data layouts, while on GPUs, it may indicate better adaptive partitioning.
   
    \item \textbf{GPU-based AMG implementations} exhibit more scalable behavior, especially at coarse levels where fewer ranks are involved, leading to reduced contention and overhead.
\end{itemize}

These findings suggest that improving AMG scalability on CPUs will require more communication optimization, while GPU implementations are already benefiting from hardware-aware, hierarchical load balancing.
\subsection{Laghos}


Laghos, a high-order Lagrangian hydrodynamics solver, performs frequent interprocess communication, particularly during mesh updates and boundary data exchanges (halo exchanges). For this benchmark, our goal was to analyze its strong scaling behavior, that is, performance under a fixed problem size with increasing process count and to provide insights based on the observed trends. Data volume per rank goes down as scale goes up since the local portion of the problem gets smaller for strong scaling.

We ran Laghos using the rs2–rp2 configuration on the Dane cluster. Here, rs2 denotes the second-level mesh refinement strategy, leading to a higher resolution computational grid, while rp2 specifies a predefined set of solver parameters, including time step size and polynomial order \cite{LAGHOS_REAL}.

\begin{figure}
    \centering
    \includegraphics[width=0.5\textwidth]{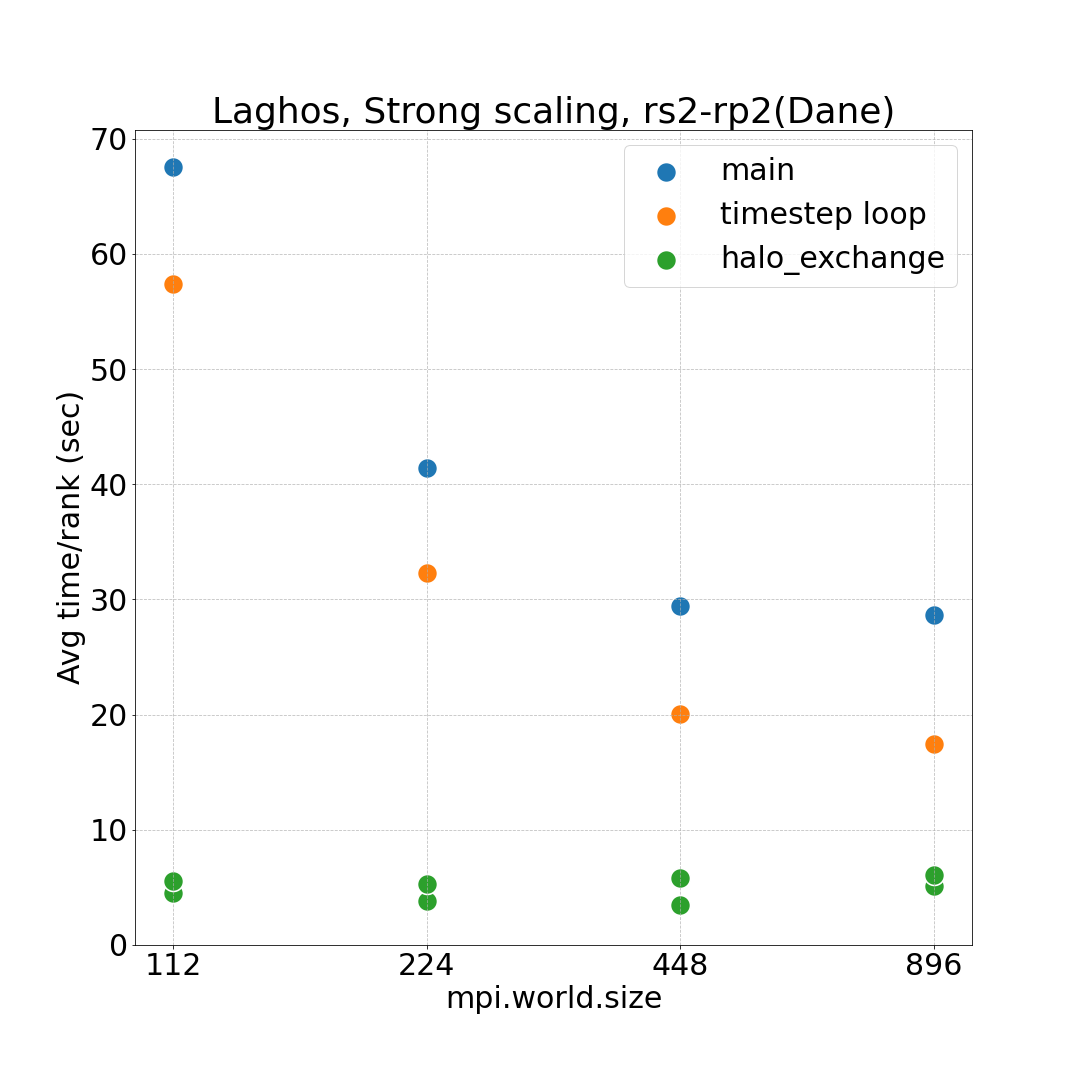}
    \caption{Average time per rank for Laghos on CPU-Dane system }   
    \label{fig:laghos_avgtimerank_laghos}
\end{figure}

Figure~\ref{fig:laghos_avgtimerank_laghos} shows the time spent in each of the annotated regions in Laghos, averaged across all ranks. The average time per rank for the main and timestep regions decreases with process count. This is because of the reduced problem size per process as we scale up. The halo\_exchange operation time varies slightly as process count increases. We observe two levels(green dots) in the plot; for the Broadcast and Reduction phases of the timestep loop. 

\section{Bandwidth and Message Rate Analysis}
\label{sec:bandwidth}

We now compare average messaging rate and bandwidth of the three benchmark applications to illustrate how each benchmark stresses the communication layer differently.
First, results for the Dane system are covered, then Tioga results are covered for all three benchmarks.

\subsection{Results on Dane}
\begin{figure*}
    \centering
    \includegraphics[width=1.0\textwidth]{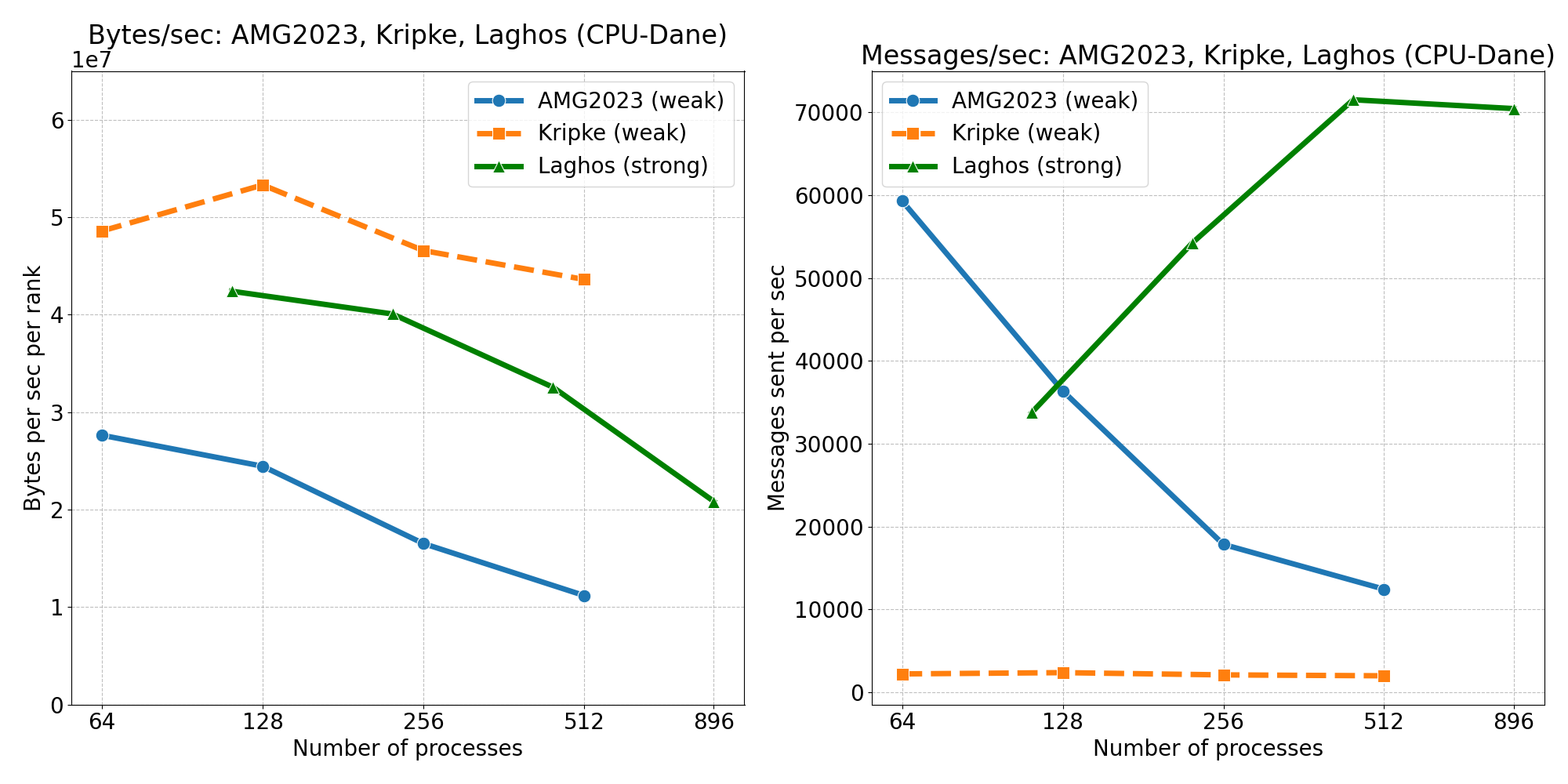}
    \caption{Bytes per second for AMG2023, Kripke and Laghos on CPU }   
    \label{fig:kripkeAMGLaghoscpu}
\end{figure*}

Figure~\ref{fig:kripkeAMGLaghoscpu} shows results from Dane. The lefthand plot shows the bandwidth and message rate for AMG2023 (blue), Kripke (orange) and Laghos (green). We observe that Kripke exhibits the highest bandwidth per process and maintains the lowest message rate compared to other benchmarks. AMG2023 demonstrates the lowest bandwidth but its message rate is higher for fewer processes and gradually declines as the number of processes increases. Laghos (strong scaling) shows moderate bandwidth while its message rate increases proportionately to the number of processes. 

For AMG2023, we observed that per-process bandwidth decreases with increase in process count. Initially, communication performance is approximately $\sim$30 million bytes per second, but this drops below $\sim$10 million bytes per second at 512 processes. This trend suggests a decline in per-rank communication throughput with increased parallelism.
This decline can be attributed to the limited memory bandwidth on CPUs compared to the GPU system, which significantly constrains the communication performance at large scales. The need for inter-node communication leads to increased latency and network congestion as messages traverse the network infrastructure.
AMG2023 in particular relies heavily on MPI communication to exchange coarse grid corrections. As the process count increases, these communication phases become slower and more dominant, reducing the overall scalability of the solver.

\
For Kripke, per-process bandwidth decreases as the number of processes increases. The performance starts near 50 million bytes/sec but gradually declines as the number of processes increases from 64 to 512.  
The decline is the result of the increasing number of process; the total available network bandwidth remains constant.  
When the process count is increased, each process gets a smaller portion of the bandwidth, leading to the downward trend of bytes transferred per process, as observed.


For  Laghos, we present results for the bandwidth of strong scaling experiments using \texttt{rs2-rp2} configuration.
We observed that from 112 to 224 processes, the system maintains high throughput, with performance peaking around 224 processes. In this regime, each process handles a relatively large partition of the mesh, which helps to keep communication localized and minimizes the number of messages and bytes per process. As a result, communication remains coarse-grained and efficient, with limited overhead from inter-process data exchange.
At 896 processes, a sharp drop in throughput is observed, indicating the onset of scalability bottlenecks on the Dane system. This decline in performance may be attributed to factors such as increased network contention and inefficient message aggregation, both of which can hinder communication efficiency as process counts grow.
This trend illustrates a key strong scaling challenge in Laghos: while computational load decreases with added processes, communication overhead does not scale down proportionally. Instead, it increases in relative cost due to the finer granularity of parallel work and increased communication. This emphasizes the need for communication-optimized strategies (e.g., topology-aware process mapping) when scaling Laghos or similar solvers to high core counts.

Figure~\ref{fig:kripkeAMGLaghoscpu}'s righthand plot shows the message rate for AMG2023, Kripke, and Laghos. For AMG2023, the number of messages sent per second drops significantly as the number of processes increases. Specifically, the messaging rate starts at approximately $\sim$60,000 messages per second at 64 processes and declines to around $\sim$10,000 messages per second at 512 processes. This trend illustrates a loss in messaging efficiency as scale increases, which could be an effect of memory and/or bandwidth bottlenecks. Kripke trends for message rate were already explained in Section~\ref{sec:Kripke}. 
%
For Laghos, the message rate increased the number of processes was increased until the message rate plateaued at 896 processes. The increase in message rate aligns with the decrease in overall message size (Table~\ref{tab:bytesAll}) and the decrease in bytes per second per rank. 



\subsection{Tioga Results}
\begin{figure*}
    \centering
    \includegraphics[width=1.0\textwidth]{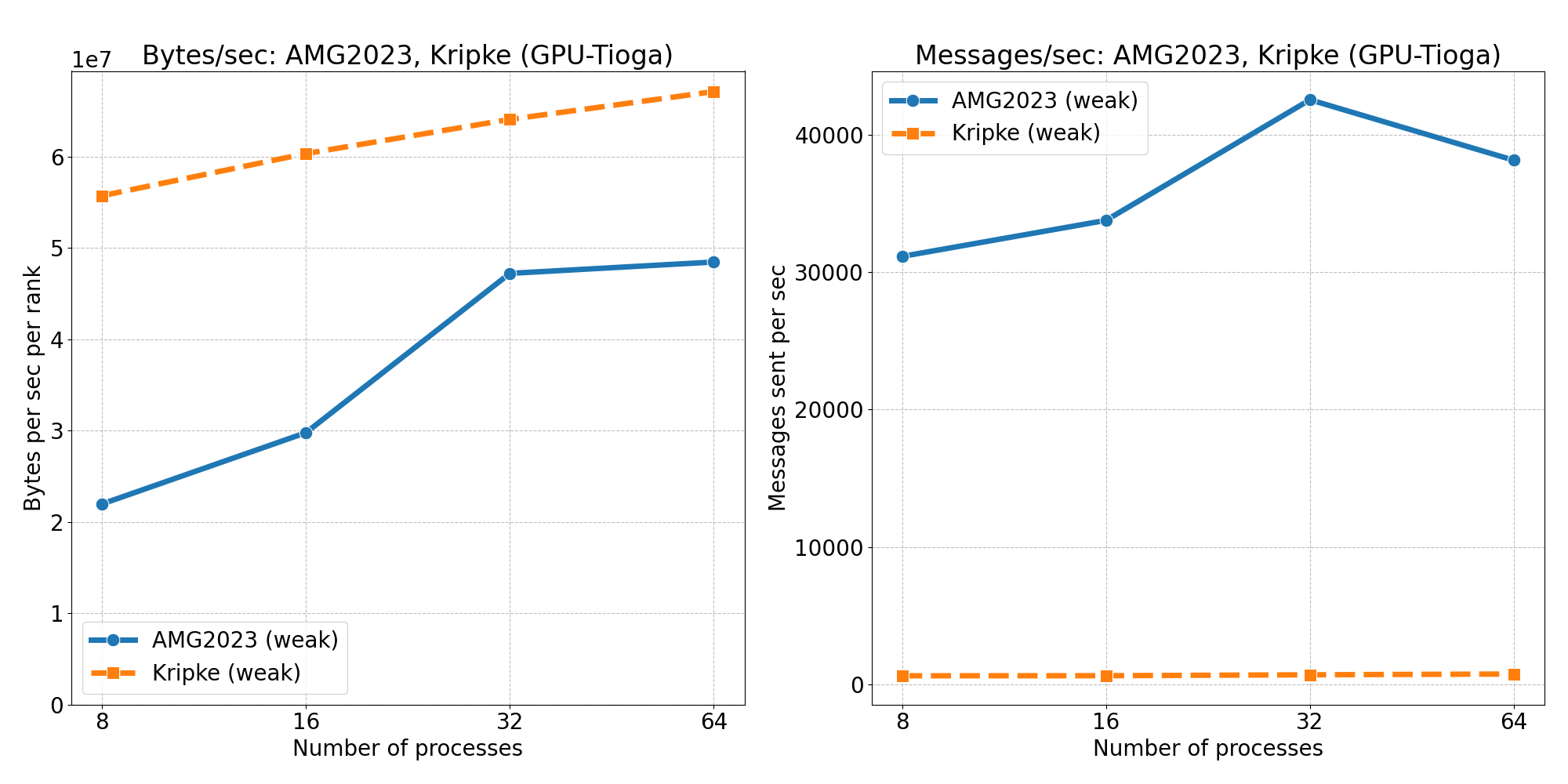}
    \caption{Bytes per second for AMG2023 and Kripke on GPU }   
    \label{fig:kripkeAmggpu}
\end{figure*}

Figure~\ref{fig:kripkeAmggpu} shows the results of the GPU-Tioga system for Kripke and AMG2023. On the left side, Kripke still shows high bandwidth and lower message rate compared to AMG2023.  The bandwidth for Kripke also increases steadily with process count on Tioga while on Dane it increases at lower process counts and declines as the process count increases. 
For Kripke on Tioga, the bytes per second per process increases from $\sim$55 million to $\sim$70 million.  
The increase in bandwidth as the number of processes increases is the result of the higher effective memory bandwidth associated with GPUs.  

For AMG2023, we observed that bandwidth increases significantly as the number of processes scales up to 32, after which the performance slightly stabilizes at 64 processes. Overall, the GPU system demonstrates high bandwidth utilization when compared to its CPU counterpart. Tioga has a high-speed memory subsystem which significantly improves the rate of data exchange between memory and processing cores. 
The AMG algorithm on GPUs benefits from efficient parallelism and fast memory access, particularly during multigrid grid transfers and interpolation steps. These operations are computationally intensive and benefit from the hardware-level acceleration provided by the GPU architecture.

Bandwidth and message rate increase with process count on Tioga, contrary to the behavior on Dane for AMG2023.
We observe better weak scaling on GPUs compared to CPU-based platforms, where communication often becomes the dominant bottleneck. The presence of high-bandwidth memory enables GPUs to sustain higher throughput at scale, enabling better weak scaling performance across increasing process counts.


In the rightmost plot of Figure~\ref{fig:kripkeAmggpu}, we observed that the number of messages per second increases from 8 to 32 processes, reaching a peak at 32 processes. However, in 64 processes, the message rate slightly decreases, indicating a change in communication behavior on larger scales.
From the system specification on Tioga, the high GPU memory bandwidth enables frequent sending of smaller messages that result into high message rates with increase in process count. 
At higher process counts, network congestion becomes more pronounced, thus the slight drop in message count at 64 processes is potentially caused by increased contention and latency in GPU-to-GPU data exchange. 

\section{Conclusion}
\label{sec:conclusion}
Through Caliper’s new \textit{communication regions} feature, this paper provided a detailed exploration of MPI communication patterns, uncovering key performance behaviors across three HPC benchmarks: AMG2023, Kripke, and Laghos. The study introduced a methodology for capturing fine-grained MPI communication metrics using annotated regions, offering a novel way to isolate and analyze communication phases within complex applications. By evaluating application-specific scaling behavior on both CPU and GPU systems, we revealed significant differences in communication efficiency, load balancing, and scalability. We also identified bottlenecks and outlined optimization opportunities tailored to diverse HPC workloads. Furthermore, this work demonstrated the value of integrating automated benchmarking via Benchpark with interactive data analysis through Thicket, enabling reproducible, scalable, and insightful performance engineering.

For application developers and performance engineers, the insights presented here not only enhance the understanding of communication behavior in large-scale systems but also inform the design of targeted optimization strategies. These include message-size tuning, communication-computation overlap, and improved data partitioning methods. Together, these contributions advance the broader goals of scalability, load balance, and efficient resource utilization, ultimately supporting the development of high-performing applications and adaptive MPI implementations in modern HPC environments.

\section*{Acknowledgements.}
\label{Acknowledgements}This work was performed under the auspices of the U\@.S\@. Department of Energy by Lawrence Livermore National Laboratory under Contract DE-AC52-07NA27344. This work was also performed with partial support from the U\@.S\@. Department of Energy's National Nuclear Security Administration (NNSA) under the Predictive Science Academic Alliance Program (PSAAP-III), Award DE-NA0003966.

\bibliographystyle{IEEEtran}
\bibliography{ref}

\end{document}